\documentclass[12pt]{article}
\usepackage{amsmath,amssymb}
\usepackage{bm}
\usepackage{mathrsfs}
\usepackage{fourier}
\usepackage{multirow}
\allowdisplaybreaks[4]
\newcommand{{\SlashD}}{D\!\!\!\!\!\!\big/}
\newcommand{{\Slashq}}{q\!\!\!\!\!\big/}
\newcommand{{\SlashF}}{{\rm F}\!\!\!\!\!/}
\usepackage[usenames]{color}

\usepackage{graphicx}

\begin{document}

\title{On diagonal representatives 
in boundary condition matrices on orbifolds}

\author{
Yoshiharu \textsc{Kawamura}\footnote{E-mail: haru@azusa.shinshu-u.ac.jp}
and Yasunari \textsc{Nishikawa}\footnote{E-mail: 17st308a@shinshu-u.ac.jp}\\
{\it Department of Physics, Shinshu University,}\\
{\it Matsumoto 390-8621, Japan}\\
}


\maketitle
\begin{abstract}
We study diagonal representatives 
of boundary condition matrices on the orbifolds $S^1/Z_2$ and $T^2/Z_m$ ($m=2, 3, 4, 6$).
We give an alternative proof of the existence of diagonal representatives 
in each equivalent class of boundary condition matrices
on $S^1/Z_2$, using a matrix exponential representation, 
and show that they do not necessarily exist 
on $T^2/Z_2$, $T^2/Z_3$, and $T^2/Z_4$.
Each equivalence class on $T^2/Z_6$ has a diagonal representative,
because its boundary conditions are determined by a single unitary matrix.
\end{abstract}

\section{Introduction}

Gauge theories defined on a higher-dimensional space-time including an orbifold
as an extra space are phenomenologically attractive, 
because gauge bosons and Higgs boson can be unified~\cite{H1,H2},
chiral fermions appear after compactification,
or Higgs mass splitting can be elegantly realized by orbifolding~\cite{K1,K2,H&N}.
Various types of models have been constructed
by using possible combinations of different choices based on
ingredients such as a structure of space-time, 
symmetries, field contents, and boundary conditions (BCs) of fields.

The study on BCs as well as dynamics is important,
because physical symmetries are determined,
in cooperation of BCs of fields and the dynamics of the Wilson line phases,
by the Hosotani mechanism~\cite{HHH&K}.
BCs are classified by equivalence relations of local gauge symmetries.
In Ref.~\cite{HH&K}, the classification of BCs has been carried out
on $S^1/Z_2$, and it is shown that each equivalence class has 
a diagonal representative, using matrix representations.
For the orbifolds $T^2/Z_m$ ($m=2, 3, 4, 6$),
it has been done, in a limited way,
for a class with a diagonal representative~\cite{KK&M,K&M,G&K},
and hence we have no definite answer 
whether each equivalence class has a diagonal representative
on the orbifolds.\footnote{
In Ref.~\cite{HN&T,K&M}, the classification of equivalence classes
has been examined in $SU(2)$ gauge theory on $T^2/Z_2$.
}

In this paper, we study diagonal representatives 
of BC matrices on the orbifolds $S^1/Z_2$ and $T^2/Z_m$ ($m=2, 3, 4, 6$).
We give an alternative proof of the existence of diagonal representatives 
in each equivalent class of BC matrices on $S^1/Z_2$, 
using a matrix exponential representation, 
and show that they do not necessarily exist 
on $T^2/Z_2$, $T^2/Z_3$, and $T^2/Z_4$.
It is shown that each equivalence class on $T^2/Z_6$ has a diagonal representative,
because its BCs are determined by a single unitary matrix.

The outline of this paper is as follows.
In the next section, we give a simple proof that there exist diagonal representatives
in all equivalence classes on $S^1/Z_2$, using a matrix exponential representation.
We study presence or absence 
of diagonal representatives on $T^2/Z_3$ in Sect. 3
and that on $T^2/Z_2$, $T^2/Z_4$, and $T^2/Z_6$ in Sect. 4.
In the last section, we give conclusions and discussions.

\section{Diagonal representatives on $S^1/Z_2$}

\subsection{Boundary conditions on $S^1/Z_2$ and equivalence classes}

The space-time is assumed to be factorized into 
a product of 4-dimensional Minkowski space-time $M^4$ and the orbifold $S^1/Z_2$.
The $S^1/Z_2$ is obtained by dividing the circle $S^1$ 
(with the identification $y\sim y+2\pi R$)
by the $Z_2$ transformation $y \rightarrow -y$.
Here, $y$ and $R$ are a coordinate and the radius of $S^1$, respectively.
The point $y$ is identified with $-y$ on $S^1/Z_2$,
and the space is regarded as an interval with length $\pi R$.
The both end points $y = 0$ and $\pi R$ are fixed points under the $Z_2$ transformation.
For $Z_2$ transformations around $y = 0$ and $\pi R$, and a loop translation along $S^1$,
each defined by
\begin{eqnarray}
R_0: y \to -y,~~ R_1: y \to 2 \pi R - y,~~ T: y \to y + 2\pi R,
\label{RT}
\end{eqnarray}
the following relations hold:
\begin{eqnarray}
R_0^2 = I,~~ R_1^2 = I,~~ T= R_1 R_0,~~ TR_0 T = R_0,~~ TR_1 T = R_1,
\label{RT-rel}
\end{eqnarray}
where $I$ is the identity operation.

Let a 5-dimensional scalar field $\phi$ be a multiplet of some transformation group $G$
concerning some internal symmetries, and
the Lagrangian density $\mathscr{L}$ be invariant under the transformation
$\phi(x, y) \to \phi'(x, y) = T_{\phi}[V] \phi(x, y)$, i.e.,
\begin{eqnarray}
\mathscr{L}(T_{\phi}[V]\phi(x, y)) = \mathscr{L}(\phi(x, y)),
\label{L-inv}
\end{eqnarray}
where $x$ is an abbreviation for a coordinate $x^{\mu}=(t, \bm{x})$ of $M^4$,
$T_{\phi}[V]$ is a representation matrix of $G$,
and $V$ is that of a fundamental representation.
From the requirement that $\mathscr{L}$
should be invariant under $R_0$, $R_1$, and $T$, i.e.,
\begin{eqnarray}
\mathscr{L}(\phi(x, -y)) = \mathscr{L}(\phi(x, 2\pi R -y)) = \mathscr{L}(\phi(x, y+2\pi R))
= \mathscr{L}(\phi(x, y)),
\label{L}
\end{eqnarray}
the BCs of $\phi(x, y)$ on $S^1/Z_2$
are determined as
\begin{eqnarray}
&~& \phi(x, -y) = \eta_{0} T_{\phi}[P_0] \phi(x, y),~~ 
\phi(x, 2\pi R - y) = \eta_{1} T_{\phi}[P_1] \phi(x, y),~~
\nonumber \\
&~& \phi(x, y + 2\pi R) = \eta_{0}\eta_{1} T_{\phi}[U] \phi(x, y),
\label{BC}
\end{eqnarray}
where $\eta_{0}$ and $\eta_{1}$ are intrinsic $Z_2$ parities
whose values are $1$ or $-1$, 
and $T_{\phi}[P_0]$, $T_{\phi}[P_1]$, and
$T_{\phi}[U]$ are representation matrices, which are elements of $G$.
The $P_0$, $P_1$, and $U$ are those of a fundamental representation, and
satisfy the same relations as Eqs.~(\ref{RT-rel}):
\begin{eqnarray}
P_0^2 = I,~~ P_1^2 = I,~~ U = P_1 P_0,~~ U P_0 U = P_0,~~ U P_1 U = P_1,
\label{PU-rel}
\end{eqnarray}
where $I$ is the unit matrix.
We refer to $P_0$, $P_1$, and $U$ as BC matrices.
The same holds for $T_{\phi}[P_0]$, $T_{\phi}[P_1]$, and $T_{\phi}[U]$.

Next, we explain equivalence classes of the BCs.
The BCs relating to a global unitary transformation are equivalent:
\begin{eqnarray}
(W^{\dagger} P_0 W,~ W^{\dagger} P_1 W,~ W^{\dagger} U W) \sim (P_0, P_1, U),
\label{equ-W}
\end{eqnarray}
where $W$ is an arbitrary global (or space-time independent) unitary matrix.
Furthermore, if the system has local gauge symmetries,
there are equivalence relations of them.
Under a specific gauge transformation 
$\phi(x,y) \to \phi'(x,y) = T_{\phi}[\varOmega] \phi(x,y)$,
the BCs of $\phi$ change as
\begin{eqnarray}
&~& \phi'(x, -y) = \eta_0 T_{\phi}[P'_0] \phi'(x, y), ~~
\phi'(x, 2 \pi R -y) = \eta_1 T_{\phi}[P'_1] \phi'(x, y), ~~
\nonumber \\
&~& \phi'(x, y + 2 \pi R) = \eta_0 \eta_1 T_{\phi}[U'] \phi'(x, y),
\label{BC'}
\end{eqnarray}
where $\varOmega = \varOmega(x,y)$ is a gauge transformation function 
and BC matrices with primes are given by
\begin{eqnarray}
&~& P'_0 = \varOmega(x,-y) P_0  \varOmega^\dagger (x,y), ~~
P'_1 = \varOmega(x, 2\pi R -y) P_1 \varOmega^\dagger (x,y), ~~
\nonumber \\
&~& U' = \varOmega(x,y+2\pi R) U \varOmega^\dagger (x,y).
\label{P'}
\end{eqnarray}
The point is that {\it the BC matrices do not necessarily 
agree with the original ones, i.e.,
$(P_0, P_1, U) \ne (P'_0, P'_1, U')$, for a singular gauge transformation.}
Because physics is invariant under gauge transformations,
the two sets of the BCs in Eqs.~(\ref{P'}) should be equivalent:
\begin{eqnarray}
(P'_0, P'_1, U') \sim (P_0, P_1, U).
\label{equ}
\end{eqnarray}
The equivalence relations (\ref{equ-W})
and (\ref{equ}) defines equivalence classes of the BCs.

Because two of $P_0$, $P_1$, and $U$ are independent,
we choose $P_0$ and $P_1$ as independent ones.
Then, $U$ is determined by $U = P_1 P_0$.
It is shown that each equivalence class has a diagonal representative for 
$P_0$ and $P_1$, using matrix representations~\cite{HH&K}.
In the next section, we give an alternative proof of the existence of
diagonal representatives, using a matrix exponential representation.

\subsection{Existence proof of diagonal representatives}

Let $P_0$ and $P_1$ be realized by $N \times N$ unitary matrices
with $N$-dimensional fundamental representation.
From $P_0^2 = I$ and $P_1^2 = I$,
we obtain the relations:
\begin{eqnarray}
P^{\dagger}_0 = P_0^{-1} = P_0,~~ P^{\dagger}_1 = P_1^{-1} = P_1,
\label{P-1=P}
\end{eqnarray}
and hence both $P_0$ and $P_1$ are unitary and hermitian matrices.
Because those matrices are, in general, 
diagonalized by global unitary transformations,
$P_0$ and $P_1$ can be written as
\begin{eqnarray}
P_0 = W_0 P_0^{\rm (D)}W^{\dagger}_0,~~
P_1 = W_1 P_1^{\rm (D)}W^{\dagger}_1 
\label{W01}
\end{eqnarray}
where $W_0$ and $W_1$ are some unitary matrices,
and $P_0^{\rm (D)}$ and $P_1^{\rm (D)}$ are diagonal matrices
whose diagonal components are $1$ or $-1$.
After rearrangement of the rows and columns,
$P_0^{\rm (D)}$ and $P_1^{\rm (D)}$ are expressed 
and specified by three non-negative integers $(p, q, r)$ such that
\begin{eqnarray}
&~& \mbox{diag} P_0^{\rm (D)} = (\overbrace{+1, \cdots, +1, +1, \cdots, +1,
 -1, \cdots, -1, -1, \cdots, -1}^N),~~
\label{diag0} \\
&~& \mbox{diag} P_1^{\rm (D)} = (\underbrace{+1, \cdots, +1}_{p}, 
\underbrace{-1, \cdots, -1}_{q},
\underbrace{+1, \cdots, +1}_r, \underbrace{-1, \cdots, -1}_{s = N-p-q-r}),
\label{diag1}
\end{eqnarray}
where $0 \leq p, q, r, s \leq N$.

Starting from arbitrary $P_0$ and $P_1$, they are transformed as
\begin{eqnarray}
P_0 \xrightarrow{W^{\dagger}_0 P_0 W_0} P_0^{\rm (D)},~~
P_1 \xrightarrow{W^{\dagger}_0 P_1 W_0} W P_1^{\rm (D)} W^{\dagger},
\label{P0W0}
\end{eqnarray}
where $W (=W^{\dagger}_0 W_1)$ is also a unitary matrix.
Then, the problem whether arbitrary $P_0$ and $P_1$ can be diagonalized simultaneously
by a global unitary transformation
and a local gauge transformation
is restated whether an arbitrary hermitian and unitary matrix 
$P_1 =W P_1^{\rm (D)} W^{\dagger}$
can be diagonalized by a global unitary transformation
and a local gauge transformation, keeping $P_0$ 
in some diagonal form $\tilde{P}_0^{\rm (D)}$ with $(\tilde{P}_0^{\rm (D)})^2 = I$.
More specifically, it is whether there are a unitary matrix $\widetilde{W}$
and a gauge transformation function $\varOmega(y)$, that satisfy the relations:
\begin{eqnarray}
&~& \varOmega(-y)\widetilde{W}^{\dagger} P_0^{\rm (D)} 
\widetilde{W} \varOmega^{\dagger}(y) = \tilde{P}_0^{\rm (D)},~~
\label{P0D}\\
&~& \varOmega(2\pi R-y)\widetilde{W}^{\dagger} (W P_1^{\rm (D)} W^{\dagger}) 
\widetilde{W} \varOmega^{\dagger}(y) = \tilde{P}_1^{\rm (D)},
\label{P1D}
\end{eqnarray}
where $x$ is omitted in $\varOmega$, 
and $\tilde{P}_1^{\rm (D)}$ is a diagonal matrix with $(\tilde{P}_1^{\rm (D)})^2 = I$.
If the answer is affirmative, it implies that each equivalence class
of BC matrices contains diagonal representatives.

In the following, it is shown that the relations 
(\ref{P0D}) and (\ref{P1D}) hold on
with $\tilde{P}_0^{\rm (D)} = P_0^{\rm (D)}$
and $\tilde{P}_1^{\rm (D)} = P_1^{\rm (D)}$,
by using the feature (see Appendix A) that an arbitrary unitary matrix $W$
can be expressed by the matrix exponential representation:
\begin{eqnarray}
W = e^{i\left(\xi^{a_{\scalebox{0.6}{$\displaystyle{++}$}}} 
T^{a_{\scalebox{0.6}{$\displaystyle{++}$}}}  
+ \xi^{a_{\scalebox{0.6}{$\displaystyle{+-}$}}} 
T^{a_{\scalebox{0.6}{$\displaystyle{+-}$}}}\right)}
e^{i \xi^{a_{\scalebox{0.6}{$\displaystyle{--}$}}} 
T^{a_{\scalebox{0.6}{$\displaystyle{--}$}}}}
e^{i\left(\eta^{a_{\scalebox{0.6}{$\displaystyle{++}$}}} 
T^{a_{\scalebox{0.6}{$\displaystyle{++}$}}}  
+ \xi^{a_{\scalebox{0.6}{$\displaystyle{-+}$}}}
T^{a_{\scalebox{0.6}{$\displaystyle{-+}$}}}\right)},
\label{W}
\end{eqnarray}
where $\xi^{a_{\scalebox{0.6}{$\displaystyle{++}$}}}$,
$\xi^{a_{\scalebox{0.6}{$\displaystyle{+-}$}}}$,
$\xi^{a_{\scalebox{0.6}{$\displaystyle{-+}$}}}$,
$\xi^{a_{\scalebox{0.6}{$\displaystyle{--}$}}}$,
and $\eta^{a_{\scalebox{0.6}{$\displaystyle{++}$}}}$ are real parameters,
$T^{a_{\scalebox{0.6}{$\displaystyle{++}$}}}$,
$T^{a_{\scalebox{0.6}{$\displaystyle{+-}$}}}$,
$T^{a_{\scalebox{0.6}{$\displaystyle{-+}$}}}$, and
$T^{a_{\scalebox{0.6}{$\displaystyle{--}$}}}$
are generators (Lie algebras) represented by hermitian matrices,
that satisfy the relations: 
\begin{eqnarray}
&~& P_0^{\rm (D)} T^{a_{\scalebox{0.6}{$\displaystyle{++}$}}}
= T^{a_{\scalebox{0.6}{$\displaystyle{++}$}}} P_0^{\rm (D)},~~
P_1^{\rm (D)} T^{a_{\scalebox{0.6}{$\displaystyle{++}$}}}
= T^{a_{\scalebox{0.6}{$\displaystyle{++}$}}} P_1^{\rm (D)},~~
\label{T++}\\
&~& P_0^{\rm (D)} T^{a_{\scalebox{0.6}{$\displaystyle{+-}$}}}
= T^{a_{\scalebox{0.6}{$\displaystyle{+-}$}}} P_0^{\rm (D)},~~
P_1^{\rm (D)} T^{a_{\scalebox{0.6}{$\displaystyle{+-}$}}}
= -T^{a_{\scalebox{0.6}{$\displaystyle{+-}$}}} P_1^{\rm (D)},~~
\label{T+-}\\
&~& P_0^{\rm (D)} T^{a_{\scalebox{0.6}{$\displaystyle{-+}$}}}
= -T^{a_{\scalebox{0.6}{$\displaystyle{-+}$}}} P_0^{\rm (D)},~~
P_1^{\rm (D)} T^{a_{\scalebox{0.6}{$\displaystyle{-+}$}}}
= T^{a_{\scalebox{0.6}{$\displaystyle{-+}$}}} P_1^{\rm (D)},~~
\label{T-+}\\
&~& P_0^{\rm (D)} T^{a_{\scalebox{0.6}{$\displaystyle{--}$}}}
= - T^{a_{\scalebox{0.6}{$\displaystyle{--}$}}} P_0^{\rm (D)},~~
P_1^{\rm (D)} T^{a_{\scalebox{0.6}{$\displaystyle{--}$}}}
= -T^{a_{\scalebox{0.6}{$\displaystyle{--}$}}} P_1^{\rm (D)}.
\label{T--}
\end{eqnarray}
In Eq.~(\ref{W}),
the summation over indices $a_{++}$, $a_{+-}$, $a_{-+}$, and $a_{--}$ are carried out.

By inserting Eq.~(\ref{W}) into $P_1 = W P_1^{\rm (D)} W^{\dagger}$
and using the relations (\ref{T++}), (\ref{T-+}), and (\ref{T--}),
$P_1$ is expressed by
\begin{eqnarray}
&~& P_1 = e^{i\left(\xi^{a_{\scalebox{0.6}{$\displaystyle{++}$}}} 
T^{a_{\scalebox{0.6}{$\displaystyle{++}$}}}  
+ \xi^{a_{\scalebox{0.6}{$\displaystyle{+-}$}}} 
T^{a_{\scalebox{0.6}{$\displaystyle{+-}$}}}\right)}
e^{i \xi^{a_{\scalebox{0.6}{$\displaystyle{--}$}}} 
T^{a_{\scalebox{0.6}{$\displaystyle{--}$}}}}
e^{i\left(\eta^{a_{\scalebox{0.6}{$\displaystyle{++}$}}} 
T^{a_{\scalebox{0.6}{$\displaystyle{++}$}}}  
+ \xi^{a_{\scalebox{0.6}{$\displaystyle{-+}$}}}
T^{a_{\scalebox{0.6}{$\displaystyle{-+}$}}}\right)} 
\nonumber \\
&~& ~~~~~~~~ \times P_1^{\rm (D)} 
e^{-i\left(\eta^{b_{\scalebox{0.6}{$\displaystyle{++}$}}} 
T^{b_{\scalebox{0.6}{$\displaystyle{++}$}}}  
+ \xi^{b_{\scalebox{0.6}{$\displaystyle{-+}$}}}
T^{b_{\scalebox{0.6}{$\displaystyle{-+}$}}}\right)}
e^{-i \xi^{b_{\scalebox{0.6}{$\displaystyle{--}$}}} 
T^{b_{\scalebox{0.6}{$\displaystyle{--}$}}}}
e^{-i\left(\xi^{b_{\scalebox{0.6}{$\displaystyle{++}$}}} 
T^{b_{\scalebox{0.6}{$\displaystyle{++}$}}}  
+ \xi^{b_{\scalebox{0.6}{$\displaystyle{+-}$}}} 
T^{b_{\scalebox{0.6}{$\displaystyle{+-}$}}}\right)}
\nonumber \\
&~& ~~~~~\! = e^{i\left(\xi^{a_{\scalebox{0.6}{$\displaystyle{++}$}}} 
T^{a_{\scalebox{0.6}{$\displaystyle{++}$}}}  
+ \xi^{a_{\scalebox{0.6}{$\displaystyle{+-}$}}} 
T^{a_{\scalebox{0.6}{$\displaystyle{+-}$}}}\right)}
e^{i \xi^{a_{\scalebox{0.6}{$\displaystyle{--}$}}} 
T^{a_{\scalebox{0.6}{$\displaystyle{--}$}}}}
 P_1^{\rm (D)} 
e^{-i \xi^{b_{\scalebox{0.6}{$\displaystyle{--}$}}} 
T^{b_{\scalebox{0.6}{$\displaystyle{--}$}}}}
e^{-i\left(\xi^{b_{\scalebox{0.6}{$\displaystyle{++}$}}} 
T^{b_{\scalebox{0.6}{$\displaystyle{++}$}}}  
+ \xi^{b_{\scalebox{0.6}{$\displaystyle{+-}$}}} 
T^{b_{\scalebox{0.6}{$\displaystyle{+-}$}}}\right)}
\nonumber \\
&~& ~~~~~\! = e^{i\left(\xi^{a_{\scalebox{0.6}{$\displaystyle{++}$}}} 
T^{a_{\scalebox{0.6}{$\displaystyle{++}$}}}  
+ \xi^{a_{\scalebox{0.6}{$\displaystyle{+-}$}}} 
T^{a_{\scalebox{0.6}{$\displaystyle{+-}$}}}\right)}
e^{2 i \xi^{a_{\scalebox{0.6}{$\displaystyle{--}$}}} 
T^{a_{\scalebox{0.6}{$\displaystyle{--}$}}}}
 P_1^{\rm (D)} 
e^{-i\left(\xi^{b_{\scalebox{0.6}{$\displaystyle{++}$}}} 
T^{b_{\scalebox{0.6}{$\displaystyle{++}$}}}  
+ \xi^{b_{\scalebox{0.6}{$\displaystyle{+-}$}}} 
T^{b_{\scalebox{0.6}{$\displaystyle{+-}$}}}\right)}.
\label{P1W}
\end{eqnarray}
Then, by using $\widetilde{W}$ and $\varOmega(y)$ defined by
\begin{eqnarray}
\widetilde{W} \equiv e^{i\left(\xi^{a_{\scalebox{0.6}{$\displaystyle{++}$}}} 
T^{a_{\scalebox{0.6}{$\displaystyle{++}$}}}  
+ \xi^{a_{\scalebox{0.6}{$\displaystyle{+-}$}}} 
T^{a_{\scalebox{0.6}{$\displaystyle{+-}$}}}\right)},~~
\varOmega(y)
\equiv e^{- \frac{i y}{\pi R} \xi^{a_{\scalebox{0.6}{$\displaystyle{--}$}}} 
T^{a_{\scalebox{0.6}{$\displaystyle{--}$}}}},
\label{WOmega}
\end{eqnarray}
we find that the left hand sides of 
the relations (\ref{P0D}) and (\ref{P1D}) 
become diagonal matrices such that
\begin{eqnarray}
&~& \varOmega(-y)\widetilde{W}^{\dagger} P_0^{\rm (D)} 
\widetilde{W} \varOmega^{\dagger}(y) 
= \varOmega(-y) P_0^{\rm (D)} \varOmega^{\dagger}(y) = P_0^{\rm (D)},~~
\label{P0D-left}\\
&~& \varOmega(2\pi R-y)\widetilde{W}^{\dagger} P_1 
\widetilde{W} \varOmega^{\dagger}(y) = 
\varOmega(2\pi R-y) e^{2 i \xi^{a_{\scalebox{0.6}{$\displaystyle{--}$}}} 
T^{a_{\scalebox{0.6}{$\displaystyle{--}$}}}} P_1^{\rm (D)} 
\varOmega^{\dagger}(y) = P_1^{\rm (D)}.
\label{P1D-left}
\end{eqnarray}
Here, we use the relations:
\begin{eqnarray}
&~& P_0^{\rm (D)} \widetilde{W} = \widetilde{W} P_0^{\rm (D)},~~
P_0^{\rm (D)} \varOmega^{\dagger}(y) = \varOmega(y)P_0^{\rm (D)},~~
P_1^{\rm (D)} \varOmega^{\dagger}(y) = \varOmega(y)P_1^{\rm (D)},
\label{POmega}\\
&~& \varOmega(-y)\varOmega(y) = I,~~
\varOmega(2\pi R-y) \varOmega(y) = e^{-2 i \xi^{a_{\scalebox{0.6}{$\displaystyle{--}$}}} 
T^{a_{\scalebox{0.6}{$\displaystyle{--}$}}}}.
\label{Omega^2}
\end{eqnarray}
This completes the proof.

\section{Diagonal representatives on $T^2/Z_3$}

\subsection{Boundary conditions on $T^2/Z_3$ and equivalence classes}

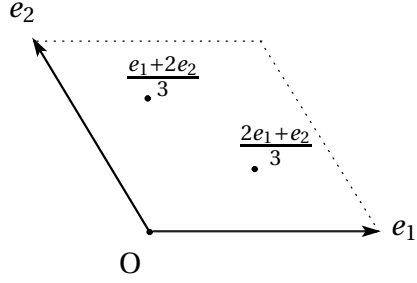
\begin{figure}[ht!]
\caption{Orbifold $T^2/Z_3$}
\label{FZ3}
\begin{center}
\unitlength 0.1in
\begin{picture}( 20.0000, 13.2000)(  8.9000,-20.3000)
%
\special{pn 13}%
\special{pa 1650 1976}%
\special{pa 2820 1976}%
\special{fp}%
\special{sh 1}%
\special{pa 2820 1976}%
\special{pa 2754 1956}%
\special{pa 2768 1976}%
\special{pa 2754 1996}%
\special{pa 2820 1976}%
\special{fp}%
%
\special{pn 13}%
\special{pa 1630 1980}%
\special{pa 1026 978}%
\special{fp}%
\special{sh 1}%
\special{pa 1026 978}%
\special{pa 1044 1046}%
\special{pa 1054 1024}%
\special{pa 1078 1026}%
\special{pa 1026 978}%
\special{fp}%
%
\special{pn 8}%
\special{pa 1040 980}%
\special{pa 2230 980}%
\special{dt 0.045}%
%
\special{pn 8}%
\special{pa 2210 980}%
\special{pa 2828 1974}%
\special{dt 0.045}%
\put(14.7000,-22.0000){\makebox(0,0)[lb]{O}}%
\put(28.9000,-20.2000){\makebox(0,0)[lb]{$e_1$}}%
\put(8.9000,-8.8000){\makebox(0,0)[lb]{$e_2$}}%
\put(15.0000,-12.7000){\makebox(0,0)[lb]{$\frac{e_1+2e_2}{3}$}}%
\put(20.9000,-16.3000){\makebox(0,0)[lb]{$\frac{2e_1+e_2}{3}$}}%
%
\special{pn 13}%
\special{sh 1}%
\special{ar 2180 1650 10 10 0  6.28318530717959E+0000}%
\special{sh 1}%
\special{ar 2180 1650 10 10 0  6.28318530717959E+0000}%
%
\special{pn 13}%
\special{sh 1}%
\special{ar 1620 1280 10 10 0  6.28318530717959E+0000}%
\special{sh 1}%
\special{ar 1620 1280 10 10 0  6.28318530717959E+0000}%
%
\special{pn 13}%
\special{sh 1}%
\special{ar 1630 1980 10 10 0  6.28318530717959E+0000}%
\special{sh 1}%
\special{ar 1630 1980 10 10 0  6.28318530717959E+0000}%
\end{picture}%
\end{center}
\end{figure}

The orbifold $T^2/Z_3$ is obtained by dividing a two-dimensional lattice $T^2$ 
(with the identification $z \sim z + e_1$ and $z \sim z + e_2$)
by the $Z_3$ transformation $z \rightarrow \omega z$ ($\omega = e^{2\pi i/3}$).
Here, $z$ is a complex coordinate of $T^2$,
and $e_1$ and $e_2$ are basis vectors of $T^2$.
The point $z$ is identified with $\omega z$ and $\bar{\omega} z$ 
($\bar{\omega} = \omega^2 = e^{4\pi i/3}$) on $T^2/Z_3$.
We take $e_1=1$ and $e_2 = \omega$, for simplicity.
Then, the points $z = 0$, $(2+\omega)/3$ and $(1+2\omega)/3$ 
are fixed points under the $Z_3$ transformation.
The resultant space is depicted in Figure \ref{FZ3}.
For $Z_3$ transformations around these fixed points and shifts along basis vectors,
each defined by
\begin{eqnarray}
&~& R_0: z \to \omega z,~~ R_1: z \to \omega z + 1,~~ R_2: z \to \omega z + 1 + \omega,~~ 
\nonumber \\
&~& T_1: z \to z + 1,~~ T_2: z \to z + \omega,
\label{RT-Z3}
\end{eqnarray}
the following relations hold:
\begin{eqnarray}
&~& R_0^3 = I,~~ R_1^3 = I,~~ R_2^3 = I,~~ R_2 R_0 R_1 = I,~~ 
R_0 R_1 R_2 = I,~~ R_1 R_2 R_0 = I,~~
\nonumber \\
&~& R_1 = T_1 R_0,~~ R_2 = T_2 T_1 R_0,~~ T_1 T_2 = T_2 T_1,
\label{RT-rel-Z3}
\end{eqnarray}
where $I$ is the identity operation.
Because two of them are independent,
we choose $R_0$ and $R_1$ as independent ones.

Let a 6-dimensional scalar field $\phi$ be a multiplet of $G$, and
$\mathscr{L}$ be invariant under
$\phi(x, z, \bar{z}) \to \phi'(x, z, \bar{z}) = T_{\phi}[V] \phi(x, z, \bar{z})$.
Here, $\bar{z}$ is the complex conjugation of $z$.
From the requirement that $\mathscr{L}$
should be a single-valued function on $T^2/Z_3$,
the BCs of $\phi(x, z, \bar{z})$ on $T^2/Z_3$
are determined as
\begin{eqnarray}
\phi(x, \omega z, \bar{\omega} \bar{z}) = \rho_{0} T_{\phi}[\Theta_0] \phi(x, z, \bar{z}),~~ 
\phi(x, \omega z + 1, \bar{\omega} \bar{z} + 1) 
= \rho_{1} T_{\phi}[\Theta_1] \phi(x, z, \bar{z}),~~
\label{BC-Z3}
\end{eqnarray}
where $\rho_{0}$ and $\rho_{1}$ are intrinsic $Z_3$ elements
whose values are $1$, $\omega$, or $\bar{\omega}$, 
and $T_{\phi}[\Theta_0]$ and $T_{\phi}[\Theta_1]$
are representation matrices.
The $\Theta_0$ and $\Theta_1$ are BC matrices 
of a fundamental representation corresponding to $R_0$ and $R_1$, and
satisfy the same relations as Eqs.~(\ref{RT-rel-Z3}):
\begin{eqnarray}
\Theta_0^3 = I,~~ \Theta_1^3 = I,
\label{Theta-rel}
\end{eqnarray}
where $I$ is the unit matrix.
The same holds for $T_{\phi}[\Theta_0]$ and $T_{\phi}[\Theta_1]$.

In the same way as the case of $S^1/Z_2$,
BCs relating to a global unitary transformation are equivalent:
\begin{eqnarray}
(W^{\dagger} \Theta_0 W,~ W^{\dagger} \Theta_1 W) \sim (\Theta_0, \Theta_1),
\label{equ-W-Z3}
\end{eqnarray}
where $W$ is an arbitrary global unitary matrix.
If the system has local gauge symmetries,
two sets of the BCs concerning a gauge transformation are also equivalent:
\begin{eqnarray}
(\Theta'_0, \Theta'_1) \sim (\Theta_0, \Theta_1),
\label{equ-Z3}
\end{eqnarray}
where $\Theta'_0$ and $\Theta'_1$ are given by
\begin{eqnarray}
\Theta'_0 
= \varOmega(x,\omega z, \bar{\omega} \bar{z}) \Theta_0  \varOmega^\dagger (x, z, \bar{z}),~~
\Theta'_1 
= \varOmega(x,\omega z + 1, \bar{\omega} \bar{z} + 1) \Theta_1  \varOmega^\dagger (x, z, \bar{z}),
\label{Theta'}
\end{eqnarray}
respectively.
Here, the BCs of $\phi$ change as
\begin{eqnarray}
\phi'(x, \omega z, \bar{\omega} \bar{z}) = \rho_0 T_{\phi}[\Theta'_0] \phi'(x, z, \bar{z}), ~~
\phi'(x, \omega z + 1, \bar{\omega} \bar{z} + 1) 
= \rho_1 T_{\phi}[\Theta'_1] \phi'(x, z, \bar{z}), 
\label{BC'-Z3}
\end{eqnarray}
under the gauge transformation 
$\phi(x,z, \bar{z}) \to \phi'(x,z, \bar{z}) = T_{\phi}[\varOmega] \phi(x,z, \bar{z})$.

\subsection{Non-existence of diagonal representatives}

Let $\Theta_0$ and $\Theta_1$ be realized by $N \times N$ unitary matrices.
Because unitary matrices are, in general, 
diagonalized by global unitary transformations,
$\Theta_0$ and $\Theta_1$ are written as
\begin{eqnarray}
\Theta_0 = W_0 \Theta_0^{\rm (D)} W^{\dagger}_0,~~
\Theta_1 = W_1 \Theta_1^{\rm (D)} W^{\dagger}_1,
\label{W01-Z3}
\end{eqnarray}
where $W_0$ and $W_1$ are some unitary matrices,
and $\Theta_0^{\rm (D)}$ and $\Theta_1^{\rm (D)}$ are diagonal matrices
whose diagonal components are $1$, $\omega$, or $\bar{\omega}$.
After rearrangement of the rows and columns,
$\Theta_0^{\rm (D)}$ and $\Theta_1^{\rm (D)}$ are expressed by
\begin{eqnarray}
&~& \mbox{diag} \Theta_0^{\rm (D)} = \left([1]_{p_1}, [1]_{p_2}, [1]_{p_3},
[\omega]_{p_4}, [\omega]_{p_5}, [\omega]_{p_6}, 
[\bar{\omega}]_{p_7}, [\bar{\omega}]_{p_8}, [\bar{\omega}]_{p_9}\right),~~
\label{diag0-Z3} \\
&~& \mbox{diag} \Theta_1^{\rm (D)} = \left([1]_{p_1}, [\omega]_{p_2}, [\bar{\omega}]_{p_3},
[1]_{p_4}, [\omega]_{p_5}, [\bar{\omega}]_{p_6}, 
[1]_{p_7}, [\omega]_{p_8}, [\bar{\omega}]_{p_9}\right),
\label{diag1-Z3}
\end{eqnarray}
where $[1]_{p_a}$, $[\omega]_{p_a}$, and $[\bar{\omega}]_{p_a}$
stand for $1$, $\omega$, and $\bar{\omega}$ for all $p_a$ elements,
and $0 \leq p_a \leq N$ ($a=1, \cdots, 9$).

Starting from arbitrary $\Theta_0$ and $\Theta_1$, they are transformed as
\begin{eqnarray}
\Theta_0 \xrightarrow{W^{\dagger}_0 \Theta_0 W_0} \Theta_0^{\rm (D)},~~
\Theta_1 \xrightarrow{W^{\dagger}_0 \Theta_1 W_0} W \Theta_1^{\rm (D)} W^{\dagger},
\label{Theta0W0}
\end{eqnarray}
where $W (=W^{\dagger}_0 W_1)$ is also a unitary matrix.
We study the problem whether a unitary matrix 
$\Theta_1 =W \Theta_1^{\rm (D)} W^{\dagger}$
can be diagonalized by a global unitary transformation
and a local gauge transformation, keeping $\Theta_0$ 
in some diagonal form $\tilde{\Theta}_0^{\rm (D)}$
with $(\tilde{\Theta}_0^{\rm (D)})^3 = I$.
If the answer is negative, it means that each equivalence class
of BCs do not necessarily contain diagonal representatives.

In the following, it is shown that 
$\Theta_1 =W \Theta_1^{\rm (D)} W^{\dagger}$
cannot necessarily be diagonalized,
by using the feature that an arbitrary unitary matrix can be written
by the matrix exponential representation:
\begin{eqnarray}
W = V_1 V_2 V_3,
\label{W-Z3}
\end{eqnarray}
where $V_1$, $V_2$, and $V_3$ are unitary matrices parameterized by
\begin{eqnarray}
&~& V_1 = e^{i\left(\xi^{a_{\scalebox{0.6}{$\displaystyle{11}$}}} 
T^{a_{\scalebox{0.6}{$\displaystyle{11}$}}}  
+ \xi^{a_{\scalebox{0.6}{$\displaystyle{1\omega}$}}} 
T^{a_{\scalebox{0.6}{$\displaystyle{1\omega}$}}}
+ \xi^{a_{\scalebox{0.6}{$\displaystyle{1\bar{\omega}}$}}} 
T^{a_{\scalebox{0.6}{$\displaystyle{1\bar{\omega}}$}}}\right)},
\label{V1-Z3}\\
&~& V_2 = e^{i \left(\xi^{a_{\scalebox{0.6}{$\displaystyle{\omega\omega}$}}} 
T^{a_{\scalebox{0.6}{$\displaystyle{\omega\omega}$}}}
+ \xi^{a_{\scalebox{0.6}{$\displaystyle{\omega\bar{\omega}}$}}} 
T^{a_{\scalebox{0.6}{$\displaystyle{\omega\bar{\omega}}$}}}
+ \xi^{a_{\scalebox{0.6}{$\displaystyle{\bar{\omega}\omega}$}}} 
T^{a_{\scalebox{0.6}{$\displaystyle{\bar{\omega}\omega}$}}}
+ \xi^{a_{\scalebox{0.6}{$\displaystyle{\bar{\omega}\bar{\omega}}$}}} 
T^{a_{\scalebox{0.6}{$\displaystyle{\bar{\omega}\bar{\omega}}$}}}\right)},
\label{V2-Z3}\\
&~& V_3 = e^{i\left(\eta^{a_{\scalebox{0.6}{$\displaystyle{11}$}}} 
T^{a_{\scalebox{0.6}{$\displaystyle{11}$}}}  
+ \xi^{a_{\scalebox{0.6}{$\displaystyle{\omega 1}$}}} 
T^{a_{\scalebox{0.6}{$\displaystyle{\omega 1}$}}}
+ \xi^{a_{\scalebox{0.6}{$\displaystyle{\bar{\omega}1}$}}} 
T^{a_{\scalebox{0.6}{$\displaystyle{\bar{\omega}1}$}}}\right)}.
\label{V3-Z3}
\end{eqnarray}
Here, 
$\xi^{a_{\scalebox{0.6}{$\displaystyle{st}$}}}$
are parameters that satisfy $\overline{\xi^{a_{\scalebox{0.6}{$\displaystyle{st}$}}}}
=\xi^{a_{\scalebox{0.6}{$\displaystyle{\bar{s}\bar{t}}$}}}$,
$\eta^{a_{\scalebox{0.6}{$\displaystyle{11}$}}}$ are redundant real ones,
and $T^{a_{\scalebox{0.6}{$\displaystyle{st}$}}}$ are generators that satisfy the relations: 
\begin{eqnarray}
&~& \left(T^{a_{\scalebox{0.6}{$\displaystyle{st}$}}}\right)^{\dagger}
= T^{a_{\scalebox{0.6}{$\displaystyle{\bar{s}\bar{t}}$}}},~~ 
\left[T^{a_{\scalebox{0.6}{$\displaystyle{st}$}}}, 
T^{b_{\scalebox{0.6}{$\displaystyle{s't'}$}}}\right]
= i f_{a_{\scalebox{0.6}{$\displaystyle{st}$}}b_{\scalebox{0.6}{$\displaystyle{s't'}$}}
c_{\scalebox{0.6}{$\displaystyle{ss'~\!tt'}$}}}
T^{c_{\scalebox{0.6}{$\displaystyle{ss'~\!tt'}$}}},
\label{Tst-com}
\end{eqnarray}
where $s$, $t$, $s'$ and $t'$ are $1$, $\omega$, or $\bar{\omega}$.
From the relations:
\begin{eqnarray}
\Theta_0^{\rm (D)} T^{a_{\scalebox{0.6}{$\displaystyle{st}$}}}
= s T^{a_{\scalebox{0.6}{$\displaystyle{st}$}}} \Theta_0^{\rm (D)},~~
\Theta_1^{\rm (D)} T^{a_{\scalebox{0.6}{$\displaystyle{st}$}}}
= t T^{a_{\scalebox{0.6}{$\displaystyle{st}$}}} \Theta_1^{\rm (D)},
\label{ThetaTst}
\end{eqnarray}
we obtain the relations:
\begin{eqnarray}
\Theta_0^{\rm (D)} V_1 = V_1 \Theta_0^{\rm (D)},~~
\Theta_1^{\rm (D)} V_3 = V_3 \Theta_1^{\rm (D)}.
\label{ThetaV=VTheta}
\end{eqnarray}

By inserting the relations (\ref{V1-Z3}), (\ref{V2-Z3}), and (\ref{V3-Z3}) 
into $\Theta_1 = W \Theta_1^{\rm (D)} W^{\dagger}$
and using the second relation in Eqs.~(\ref{ThetaV=VTheta}),
$\Theta_1$ is expressed by
\begin{eqnarray}
\Theta_1 = V_1 V_2 V_3 \Theta_1^{\rm (D)} V^{\dagger}_3 V^{\dagger}_2 V^{\dagger}_1
= V_1 V_2 \Theta_1^{\rm (D)}V^{\dagger}_2 V^{\dagger}_1.
\label{Theta1W}
\end{eqnarray}
Then, $\Theta_0^{\rm (D)}$ 
and $\Theta_1 (=V_1 V_2 \Theta_1^{\rm (D)}V^{\dagger}_2 V^{\dagger}_1)$
are transformed as
\begin{eqnarray}
\Theta_0^{\rm (D)} \xrightarrow{V^{\dagger}_1 \Theta_0^{\rm (D)} V_1} \Theta_0^{\rm (D)},~~
\Theta_1 \xrightarrow{V^{\dagger}_1 \Theta_1 V_1} V_2 \Theta_1^{\rm (D)} V^{\dagger}_2,
\label{Theta0V1}
\end{eqnarray}
by using the first relation in Eqs,~(\ref{ThetaV=VTheta}).
We cannot make $V_2 \Theta_1^{\rm (D)} V^{\dagger}_2$ a diagonal matrix,
keeping $\Theta_0^{\rm (D)}$ in a diagonal form, by using a global unitary transformation,
because $V^{\dagger}_2 \Theta_0^{\rm (D)} V_2$ is not diagonal 
unless $V_2$ equals to $e^{i\theta} I$ ($\theta$:a real constant).

We investigate whether $V_2 \Theta_1^{\rm (D)} V^{\dagger}_2$ becomes a diagonal matrix,
keeping $\Theta_0^{\rm (D)}$ in a diagonal form, by using a local gauge transformation.
More specifically,
the problem is whether there is a gauge transformation function 
$\varOmega(z, \bar{z}) = e^{-i\lambda^{a_{\scalebox{0.6}{$\displaystyle{st}$}}}(z, \bar{z})
T^{a_{\scalebox{0.6}{$\displaystyle{st}$}}}}$, that satisfy the relations:
\begin{eqnarray}
&~& \varOmega(\omega z, \bar{\omega} \bar{z}) \Theta_0^{\rm (D)} 
\varOmega^{\dagger}(z, \bar{z}) = \tilde{\Theta}_0^{\rm (D)},~~
\label{Theta0D-gauge}\\
&~& \varOmega(\omega z + 1, \bar{\omega} \bar{z} + 1) 
V_2 \Theta_1^{\rm (D)} V^{\dagger}_2
\varOmega^{\dagger}(z, \bar{z}) = \tilde{\Theta}_1^{\rm (D)},
\label{Theta1D-gauge}
\end{eqnarray}
where $x$ is omitted in $\varOmega$, 
$\lambda^{a_{\scalebox{0.6}{$\displaystyle{st}$}}}(z, \bar{z})$
are functions that satisfy $\overline{\lambda^{a_{\scalebox{0.6}{$\displaystyle{st}$}}}}
=\lambda^{a_{\scalebox{0.6}{$\displaystyle{\bar{s}\bar{t}}$}}}$,
and $\tilde{\Theta}_0^{\rm (D)}$ and $\tilde{\Theta}_1^{\rm (D)}$ 
are some diagonal matrices with 
$(\tilde{\Theta}_0^{\rm (D)})^3 = I$ and $(\tilde{\Theta}_1^{\rm (D)})^3 = I$.
Under the gauge transformation, $\Theta_0^{\rm (D)}$ becomes as
\begin{eqnarray}
{\Theta'_0}^{\rm (D)} = \varOmega(\omega z, \bar{\omega} \bar{z})\Theta_0^{\rm (D)} 
\varOmega^{\dagger}(z, \bar{z}) = \varOmega(\omega z, \bar{\omega} \bar{z})
\widehat{\varOmega}^{\dagger}(z, \bar{z}) \Theta_0^{\rm (D)},
\label{Theta0D-gauge-Omega}
\end{eqnarray}
where $\widehat{\varOmega}^{\dagger}(z, \bar{z})$ is give by
\begin{eqnarray}
\widehat{\varOmega}^{\dagger}(z, \bar{z}) 
= e^{is \lambda^{a_{\scalebox{0.6}{$\displaystyle{st}$}}}(z, \bar{z})
T^{a_{\scalebox{0.6}{$\displaystyle{st}$}}}}.
\label{hatOmega-Z3}
\end{eqnarray}
By using Eq.~(\ref{hatOmega-Z3}), 
$\varOmega(\omega z, \bar{\omega} \bar{z})\widehat{\varOmega}^{\dagger}(z, \bar{z})$
is written as
\begin{eqnarray}
\varOmega(\omega z, \bar{\omega} \bar{z})\widehat{\varOmega}^{\dagger}(z, \bar{z})
= e^{-i\lambda^{a_{\scalebox{0.6}{$\displaystyle{st}$}}}(\omega z, \bar{\omega}\bar{z})
T^{a_{\scalebox{0.6}{$\displaystyle{st}$}}}}
e^{is' \lambda^{b_{\scalebox{0.6}{$\displaystyle{s't'}$}}}(z, \bar{z})
T^{b_{\scalebox{0.6}{$\displaystyle{s't'}$}}}}.
\label{Omega*hatOmega-Z3}
\end{eqnarray}
When $\varOmega(\omega z, \bar{\omega} \bar{z})\widehat{\varOmega}^{\dagger}(z, \bar{z})$
is a diagonal form, it is a unit matrix 
and $\varOmega(z, \bar{z})$ is restricted as
\begin{eqnarray}
\varOmega(z, \bar{z}) 
= e^{-i\left(z \lambda_0^{a_{\scalebox{0.6}{$\displaystyle{\omega \omega}$}}}
T^{a_{\scalebox{0.6}{$\displaystyle{\omega \omega}$}}}
+z \lambda_0^{a_{\scalebox{0.6}{$\displaystyle{\omega \bar{\omega}}$}}}
T^{a_{\scalebox{0.6}{$\displaystyle{\omega \bar{\omega}}$}}}
+\bar{z} \lambda_0^{a_{\scalebox{0.6}{$\displaystyle{\bar{\omega} \omega}$}}}
T^{a_{\scalebox{0.6}{$\displaystyle{\bar{\omega} \omega}$}}}
+\bar{z} \lambda_0^{a_{\scalebox{0.6}{$\displaystyle{\bar{\omega} \bar{\omega}}$}}}
T^{a_{\scalebox{0.6}{$\displaystyle{\bar{\omega} \bar{\omega}}$}}}
\right)},
\label{Omega-Z3-res}
\end{eqnarray}
where $\lambda_0^{a_{\scalebox{0.6}{$\displaystyle{\omega \omega}$}}}$,
$\lambda_0^{a_{\scalebox{0.6}{$\displaystyle{\omega \bar{\omega}}$}}}$,
$\lambda_0^{a_{\scalebox{0.6}{$\displaystyle{\bar{\omega} \omega}$}}}$,
and $\lambda_0^{a_{\scalebox{0.6}{$\displaystyle{\bar{\omega} \bar{\omega}}$}}}$
are some constants.

The remaining task is to examine whether 
the relation (\ref{Theta1D-gauge}) holds on or not,
with assistance of the gauge transformation function (\ref{Omega-Z3-res}).
By inserting Eqs.~(\ref{V2-Z3}) and (\ref{Omega-Z3-res}) 
into the left hand side of the relation (\ref{Theta1D-gauge}),
it is calculated as
\begin{eqnarray}
&~& \varOmega(\omega z + 1, \bar{\omega} \bar{z} + 1) 
V_2 \Theta_1^{\rm (D)} V^{\dagger}_2
\varOmega^{\dagger}(z, \bar{z}) 
\nonumber \\
&~& ~~  
= e^{-i\left\{(\omega z + 1)\lambda_0^{a_{\scalebox{0.6}{$\displaystyle{\omega \omega}$}}}
T^{a_{\scalebox{0.6}{$\displaystyle{\omega \omega}$}}}
+(\omega z + 1)\lambda_0^{a_{\scalebox{0.6}{$\displaystyle{\omega \bar{\omega}}$}}}
T^{a_{\scalebox{0.6}{$\displaystyle{\omega \bar{\omega}}$}}}
+(\bar{\omega} \bar{z} + 1)\lambda_0^{a_{\scalebox{0.6}{$\displaystyle{\bar{\omega} \omega}$}}}
T^{a_{\scalebox{0.6}{$\displaystyle{\bar{\omega} \omega}$}}}
+(\bar{\omega} \bar{z} + 1)
\lambda_0^{a_{\scalebox{0.6}{$\displaystyle{\bar{\omega} \bar{\omega}}$}}}
T^{a_{\scalebox{0.6}{$\displaystyle{\bar{\omega} \bar{\omega}}$}}}
\right\}}
\nonumber \\
&~& ~~~~~~ \times  
e^{i \left(\xi^{b_{\scalebox{0.6}{$\displaystyle{\omega\omega}$}}} 
T^{b_{\scalebox{0.6}{$\displaystyle{\omega\omega}$}}}
+ \xi^{b_{\scalebox{0.6}{$\displaystyle{\omega\bar{\omega}}$}}} 
T^{b_{\scalebox{0.6}{$\displaystyle{\omega\bar{\omega}}$}}}
+ \xi^{b_{\scalebox{0.6}{$\displaystyle{\bar{\omega}\omega}$}}} 
T^{b_{\scalebox{0.6}{$\displaystyle{\bar{\omega}\omega}$}}}
+ \xi^{b_{\scalebox{0.6}{$\displaystyle{\bar{\omega}\bar{\omega}}$}}} 
T^{b_{\scalebox{0.6}{$\displaystyle{\bar{\omega}\bar{\omega}}$}}}\right)}
\nonumber \\
&~& ~~~~~~ \times
e^{-i \left(\omega \xi^{c_{\scalebox{0.6}{$\displaystyle{\omega\omega}$}}} 
T^{c_{\scalebox{0.6}{$\displaystyle{\omega\omega}$}}}
+ \bar{\omega} \xi^{c_{\scalebox{0.6}{$\displaystyle{\omega\bar{\omega}}$}}} 
T^{c_{\scalebox{0.6}{$\displaystyle{\omega\bar{\omega}}$}}}
+ \omega \xi^{c_{\scalebox{0.6}{$\displaystyle{\bar{\omega}\omega}$}}} 
T^{c_{\scalebox{0.6}{$\displaystyle{\bar{\omega}\omega}$}}}
+ \bar{\omega} \xi^{c_{\scalebox{0.6}{$\displaystyle{\bar{\omega}\bar{\omega}}$}}} 
T^{c_{\scalebox{0.6}{$\displaystyle{\bar{\omega}\bar{\omega}}$}}}\right)}
\nonumber \\
&~& ~~~~~~ \times
e^{i\left(\omega z \lambda_0^{d_{\scalebox{0.6}{$\displaystyle{\omega \omega}$}}}
T^{d_{\scalebox{0.6}{$\displaystyle{\omega \omega}$}}}
+ \bar{\omega} z \lambda_0^{d_{\scalebox{0.6}{$\displaystyle{\omega \bar{\omega}}$}}}
T^{d_{\scalebox{0.6}{$\displaystyle{\omega \bar{\omega}}$}}}
+ \omega \bar{z} \lambda_0^{d_{\scalebox{0.6}{$\displaystyle{\bar{\omega} \omega}$}}}
T^{d_{\scalebox{0.6}{$\displaystyle{\bar{\omega} \omega}$}}}
+ \bar{\omega} \bar{z} \lambda_0^{d_{\scalebox{0.6}{$\displaystyle{\bar{\omega} \bar{\omega}}$}}}
T^{d_{\scalebox{0.6}{$\displaystyle{\bar{\omega} \bar{\omega}}$}}}
\right)} \Theta_1^{\rm (D)},
\label{Theta1D-gauge-Omega}
\end{eqnarray}
where we use the second relation of Eqs.~(\ref{ThetaTst}).
The right hand side of Eq.~(\ref{Theta1D-gauge-Omega})
is not diagonal except for a special case.
For instance, in a case with a specific $V_2$ that satisfy the conditions:
\begin{eqnarray}
\xi^{a_{\scalebox{0.6}{$\displaystyle{\omega\bar{\omega}}$}}} = 0,~~
\xi^{a_{\scalebox{0.6}{$\displaystyle{\bar{\omega}\omega}$}}} = 0,~~
\left[\xi^{a_{\scalebox{0.6}{$\displaystyle{\omega\omega}$}}} 
T^{a_{\scalebox{0.6}{$\displaystyle{\omega\omega}$}}},~
\xi^{b_{\scalebox{0.6}{$\displaystyle{\bar{\omega}\bar{\omega}}$}}} 
T^{b_{\scalebox{0.6}{$\displaystyle{\bar{\omega}\bar{\omega}}$}}}\right] = 0,
\label{Cs}
\end{eqnarray}
$\varOmega(\omega z + 1, \bar{\omega} \bar{z} + 1) 
V_2 \Theta_1^{\rm (D)} V^{\dagger}_2
\varOmega^{\dagger}(z, \bar{z})$ becomes 
the diagonal one $\Theta_1^{\rm (D)}$
by taking $\lambda_0^{a_{\scalebox{0.6}{$\displaystyle{\omega \bar{\omega}}$}}} = 0$,
$\lambda_0^{a_{\scalebox{0.6}{$\displaystyle{\bar{\omega}\omega}$}}} = 0$,
$\lambda_0^{a_{\scalebox{0.6}{$\displaystyle{\omega \omega}$}}}
= (1 - \omega) \xi^{a_{\scalebox{0.6}{$\displaystyle{\omega \omega}$}}}$,
and $\lambda_0^{a_{\scalebox{0.6}{$\displaystyle{\bar{\omega} \bar{\omega}}$}}}
= (1 - \bar{\omega}) \xi^{a_{\scalebox{0.6}{$\displaystyle{\bar{\omega} \bar{\omega}}$}}}$.
Hence, diagonal representatives do not necessarily exist 
in the equivalence classes of BC matrices on $T^2/Z_3$.

\section{Diagonal representatives on $T^2/Z_2$, $T^2/Z_4$, and $T^2/Z_6$}

First, we list basis vectors, the independent transformations relating to
identifications of points on $T^2/Z_m$ ($m=2, 3, 4, 6$),
and the corresponding BC matrices 
for the fundamental representation,
in Table \ref{character}~\cite{G&K}.
\begin{table}[htb]
\caption{The characters of $T^2/Z_m$.}
\label{character}
\begin{center}
\begin{tabular}{c|c|c|c} \hline
$T^2/Z_m$ & Basis vectors & Transformations 
& BC matrices \\ \hline\hline
$T^2/Z_2$ & $1, i$  
& $z \to -z,~ z \to 1-z,~ z \to i-z$
& $P_0,~ P_1,~ P_2$ \\ \hline
$T^2/Z_3$ & $1, e^{2\pi i/3}$  
& $z \to e^{2\pi i/3} z,~ z \to e^{2\pi i/3} z + 1$ 
& $\Theta_0,~ \Theta_1$ \\ \hline
$T^2/Z_4$ & $1, i$ & $z \to iz,~ z \to iz + 1$ & $\Xi_0,~ \Xi_1$ \\ \hline
$T^2/Z_6$ & $1, (-3+i\sqrt{3})/2$  
& $z \to e^{\pi i/3} z$ & $\Phi$ \\ \hline
\end{tabular}
\end{center}
\end{table}
Here, those of $T^2/Z_3$ are given, for the sake of completeness.  

\subsection{$T^2/Z_2$}
\label{app-A1}

The orbifold $T^2/Z_2$ is obtained by identifying $z + e_1$, $z + e_2$,
and $-z$ with $z$.
We take $e_1 = 1$ and $e_2 = i$.
The resultant space is depicted in Figure \ref{FZ2}.
\begin{figure}[ht!]
\caption{Orbifold $T^2/Z_2$}
\label{FZ2}
\begin{center}
\unitlength 0.1in
\begin{picture}( 15.1000, 15.7000)(  3.6000,-17.5000)
%
\special{pn 13}%
\special{pa 620 1616}%
\special{pa 1790 1616}%
\special{fp}%
\special{sh 1}%
\special{pa 1790 1616}%
\special{pa 1724 1596}%
\special{pa 1738 1616}%
\special{pa 1724 1636}%
\special{pa 1790 1616}%
\special{fp}%
%
\special{pn 13}%
\special{pa 600 1620}%
\special{pa 614 450}%
\special{fp}%
\special{sh 1}%
\special{pa 614 450}%
\special{pa 592 516}%
\special{pa 612 504}%
\special{pa 632 518}%
\special{pa 614 450}%
\special{fp}%
%
\special{pn 8}%
\special{pa 600 450}%
\special{pa 1790 450}%
\special{dt 0.045}%
%
\special{pn 8}%
\special{pa 1790 450}%
\special{pa 1790 1620}%
\special{dt 0.045}%
%
\special{pn 13}%
\special{sh 1}%
\special{ar 610 1620 10 10 0  6.28318530717959E+0000}%
\special{sh 1}%
\special{ar 600 1620 10 10 0  6.28318530717959E+0000}%
%
\special{pn 13}%
\special{sh 1}%
\special{ar 620 1010 10 10 0  6.28318530717959E+0000}%
\special{sh 1}%
\special{ar 610 1010 10 10 0  6.28318530717959E+0000}%
%
\special{pn 13}%
\special{sh 1}%
\special{ar 1190 1010 10 10 0  6.28318530717959E+0000}%
\special{sh 1}%
\special{ar 1180 1010 10 10 0  6.28318530717959E+0000}%
%
\special{pn 13}%
\special{sh 1}%
\special{ar 1200 1620 10 10 0  6.28318530717959E+0000}%
\special{sh 1}%
\special{ar 1190 1620 10 10 0  6.28318530717959E+0000}%
\put(4.4000,-18.4000){\makebox(0,0)[lb]{O}}%
\put(18.7000,-16.9000){\makebox(0,0)[lb]{$e_1$}}%
\put(5.3000,-3.5000){\makebox(0,0)[lb]{$e_2$}}%
\put(3.6000,-11.0000){\makebox(0,0)[lb]{$\frac{e_2}{2}$}}%
\put(11.3000,-19.2000){\makebox(0,0)[lb]{$\frac{e_1}{2}$}}%
\put(11.9000,-9.4000){\makebox(0,0)[lb]{$\frac{e_1+e_2}{2}$}}%
\end{picture}%
\end{center}
\end{figure}
There are four fixed points $z = 0$, $e_1/2$, $e_2/2$, $(e_1+e_2)/ 2$
under the $Z_2$ transformation $z \to -z$.
Around these points, we define six kinds of transformations:
\begin{eqnarray}
&~& R_0: z\rightarrow -z,~~
R_1: z\rightarrow e_1 - z, ~~
R_2: z\rightarrow e_2 - z, ~~
R_3: z\rightarrow e_1+e_2 - z, ~~
\nonumber \\
&~& T_1: z\rightarrow z+e_1,~~
T_2: z\rightarrow z+e_2,
\label{Z2-transf}
\end{eqnarray}
and they satisfy the relations:
\begin{eqnarray}
&~& R_0^2=I,~~R_1^2=I,~~ R_2^2=I,~~ R_3^2=I,~~ R_1=T_1 R_0,~~ R_2=T_2 R_0,
\nonumber \\
&~& R_3=T_1 T_2 R_0=R_1 R_0 R_2=R_2 R_0 R_1,~~ T_1 T_2=T_2 T_1,
\label{Z2-relations}
\end{eqnarray}
where $I$ is the identity operation.

The BC matrices satisfy the relations: 
\begin{eqnarray}
&~& P_{0}^2= I,~~ P_{1}^2=I,~~ P_{2}^2=I,~~ P_3^2=I,~~
P_{1}=T_1 P_{0},~~ P_{2}=T_2 P_{0},
\nonumber \\
&~& P_3=T_1 T_2 P_0=P_{1} P_0 P_{2}
=P_{2} P_0 P_{1},~~ T_1 T_2=T_2 T_1,
\label{Z2-Rel}
\end{eqnarray}
as the consistency conditions.
From the relations (\ref{Z2-relations}) and (\ref{Z2-Rel}),
we find that any three transformations are independent
and others are constructed as combinations of them.
We choose the transformations $R_0:z \to -z$, $R_1:z \to 1-z$
and $R_2:z \to i-z$ as independent ones.

Starting from arbitrary $P_0$, $P_1$, and $P_2$
and using a suitable unitary matrix $W_0$, they are transformed as
\begin{eqnarray}
&~& P_0 \xrightarrow{W^{\dagger}_0 P_0 W_0} P_0^{\rm (D)},~~
P_1 \xrightarrow{W^{\dagger}_0 P_1 W_0} 
\tilde{P}_1 \equiv e^{2i\left(\xi^{a_{\scalebox{0.6}{$\displaystyle{---}$}}} 
T^{a_{\scalebox{0.6}{$\displaystyle{---}$}}}  
+ \xi^{a_{\scalebox{0.6}{$\displaystyle{--+}$}}} 
T^{a_{\scalebox{0.6}{$\displaystyle{--+}$}}}\right)}
P_1^{\rm (D)},
\nonumber \\
&~& P_2 \xrightarrow{W^{\dagger}_0 P_2 W_0} 
\tilde{P}_2 \equiv V e^{2i\left(\zeta^{a_{\scalebox{0.6}{$\displaystyle{---}$}}} 
T^{a_{\scalebox{0.6}{$\displaystyle{---}$}}}  
+ \zeta^{a_{\scalebox{0.6}{$\displaystyle{-+-}$}}} 
T^{a_{\scalebox{0.6}{$\displaystyle{-+-}$}}}\right)}
P_1^{\rm (D)} V^{\dagger},
\label{P012}
\end{eqnarray}
where $V$ is a unitary matrices parameterized by
\begin{eqnarray}
V = e^{i\left(\xi^{a_{\scalebox{0.6}{$\displaystyle{+++}$}}} 
T^{a_{\scalebox{0.6}{$\displaystyle{+++}$}}}  
+ \xi^{a_{\scalebox{0.6}{$\displaystyle{++-}$}}} 
T^{a_{\scalebox{0.6}{$\displaystyle{++-}$}}}
+ \xi^{a_{\scalebox{0.6}{$\displaystyle{+-+}$}}} 
T^{a_{\scalebox{0.6}{$\displaystyle{+-+}$}}}
+\xi^{a_{\scalebox{0.6}{$\displaystyle{+--}$}}} 
T^{a_{\scalebox{0.6}{$\displaystyle{+--}$}}}\right)},
\label{V-Z2}
\end{eqnarray}
and $P_0^{\rm (D)}$, $P_1^{\rm (D)}$, and $P_2^{\rm (D)}$ are
diagonal matrices expressed by
\begin{eqnarray}
&~& \mbox{diag} P_0^{\rm (D)} = \left([1]_{p_1}, [1]_{p_2}, [1]_{p_3},
[1]_{p_4}, [-1]_{p_5}, [-1]_{p_6}, 
[-1]_{p_7}, [-1]_{p_8}\right),~~
\label{diag0-Z2} \\
&~& \mbox{diag} P_1^{\rm (D)} = \left([1]_{p_1}, [1]_{p_2}, [-1]_{p_3},
[-1]_{p_4}, [1]_{p_5}, [1]_{p_6}, [-1]_{p_7}, [-1]_{p_8}\right),
\label{diag1-Z2} \\
&~& \mbox{diag} P_2^{\rm (D)} = \left([1]_{p_1}, [-1]_{p_2}, [1]_{p_3},
[-1]_{p_4}, [1]_{p_5}, [-1]_{p_6}, [1]_{p_7}, [-1]_{p_8}\right).
\label{diag2-Z2}
\end{eqnarray}
Here, $[1]_{p_a}$ and $[-1]_{p_a}$ 
represent $1$ and $-1$ for all $p_a$ elements,
and $0 \leq p_a \leq N$ ($a=1, \cdots, 8$).
The $T^{a_{\scalebox{0.6}{$\displaystyle{stu}$}}}$  
are generators that satisfy the relations:
\begin{eqnarray}
P_0^{\rm (D)} T^{a_{\scalebox{0.6}{$\displaystyle{stu}$}}}
= s T^{a_{\scalebox{0.6}{$\displaystyle{stu}$}}} P_0^{\rm (D)},~~
P_1^{\rm (D)} T^{a_{\scalebox{0.6}{$\displaystyle{stu}$}}}
= t T^{a_{\scalebox{0.6}{$\displaystyle{stu}$}}} P_1^{\rm (D)},~~
P_2^{\rm (D)} T^{a_{\scalebox{0.6}{$\displaystyle{stu}$}}}
= u T^{a_{\scalebox{0.6}{$\displaystyle{stu}$}}} P_2^{\rm (D)},
\label{Rel-Z2}
\end{eqnarray}
where $s$, $t$, and $u$ are $+$ or $-$.

Under a local gauge transformation,
the BC matrices $P_0^{\rm (D)}$, $\tilde{P}_1$, and $\tilde{P}_2$ change as
\begin{eqnarray}
&~& {P'}_0^{\rm (D)} = \varOmega(-z, -\bar{z}) P_0^{\rm (D)}  \varOmega^\dagger (z, \bar{z}),~~
\tilde{P}'_1 = \varOmega(1 - z, 1- \bar{z}) \tilde{P}_1 \varOmega^\dagger (z, \bar{z}),~~
\nonumber \\
&~& \tilde{P}'_2 = \varOmega(i - z, -i- \bar{z}) \tilde{P}_2 \varOmega^\dagger (z, \bar{z}),
\label{P'-Z2}
\end{eqnarray}
where $\varOmega$ is a gauge transformation function.
Although $P_0^{\rm (D)}$ is invariant under the gauge transformation with
the gauge transformation function:
\begin{eqnarray}
&~& \varOmega(z, \bar{z}) 
= e^{i\left(\lambda^{a_{\scalebox{0.6}{$\displaystyle{---}$}}}(z,\bar{z})
T^{a_{\scalebox{0.6}{$\displaystyle{---}$}}}  
+ \lambda^{a_{\scalebox{0.6}{$\displaystyle{--+}$}}}(z,\bar{z}) 
T^{a_{\scalebox{0.6}{$\displaystyle{--+}$}}}
+ \lambda^{a_{\scalebox{0.6}{$\displaystyle{-+-}$}}}(z,\bar{z}) 
T^{a_{\scalebox{0.6}{$\displaystyle{-+-}$}}}\right)},
\label{Omega-Z2}\\
&~& \lambda^{a_{\scalebox{0.6}{$\displaystyle{---}$}}}(-z,-\bar{z})
= -\lambda^{a_{\scalebox{0.6}{$\displaystyle{---}$}}}(z,\bar{z}),~~
\lambda^{a_{\scalebox{0.6}{$\displaystyle{--+}$}}}(-z,-\bar{z})
= -\lambda^{a_{\scalebox{0.6}{$\displaystyle{--+}$}}}(z,\bar{z}),~~
\nonumber \\
&~& \lambda^{a_{\scalebox{0.6}{$\displaystyle{-+-}$}}}(-z,-\bar{z})
= -\lambda^{a_{\scalebox{0.6}{$\displaystyle{-+-}$}}}(z,\bar{z}),
\label{lambda-Z2}
\end{eqnarray}
local gauge transformations that make $P_0^{\rm (D)}$, $\tilde{P}_1$, and $\tilde{P}_2$
diagonal matrices simultaneously do not exist except for a special case.
Hence, diagonal representatives do not necessarily exist 
in equivalence classes of BC matrices on $T^2/Z_2$.

\subsection{$T^2/Z_4$}
\label{app-A3}

The orbifold $T^2/Z_4$ is obtained by identifying 
$z +e_1$, $z + e_2$, $iz$, and $-z$ with $z$.
We take $e_1 = 1$ and $e_2 = i$.
The resultant space is depicted as the same figure as $T^2/Z_2$.
There are two fixed points $z = 0$ 
and $(e_1+e_2)/2$ under the $Z_4$ transformation $z \to iz$
and four fixed points $z = 0$, $e_1/2$, $e_2/2$, and $(e_1+e_2)/2$ 
under the $Z_2$ transformation $z \to -z$.
Around these points, we define eight kinds of transformations:
\begin{eqnarray}
&~& R_0: z\rightarrow iz, ~~
R_1: z\rightarrow iz+e_1, ~~ R_{20}: z\rightarrow -z, 
\nonumber \\
&~& R_{21}: z\rightarrow e_1 - z,~~
R_{22}: z\rightarrow e_2 - z, ~~
R_{23}: z\rightarrow e_1+e_2 - z, 
\nonumber \\
&~& T_1: z\rightarrow z+e_1, ~~
T_2: z\rightarrow z+e_2,
\label{Z4-transf}
\end{eqnarray}
and they satisfy the relations:
\begin{eqnarray}
&~& R_0^4= I,~~ R_1^4= I,~~ R_{20}^2= I,~~ R_{21}^2= I,~~ R_{22}^2=I,~~ R_{23}^2=I,~~
R_1=T_1 R_0,~~ R_{21}=T_1 R_{20}, 
\nonumber \\
&~& R_{22}=T_2 R_{20},~~ R_{20}=R_0^2,~~ R_{21}=R_1 R_0,~~ R_{22}= R_0 R_1, 
\nonumber \\
&~& R_{23}=T_1 T_2 R_{20}=R_{21}R_{20}R_{22}=R_{22}R_{20}R_{21},~~
T_1 T_2=T_2 T_1.
\label{Z4-relations}
\end{eqnarray}
The $Z_4$ transformations $R_0$ and $R_1$ are independent of each other
and the corresponding BC matrices are denoted as $\Xi_0$ and $\Xi_1$, respectively.
Other representation matrices are determined uniquely,
if $\Xi_0$ and $\Xi_1$ are given.

Starting from arbitrary unitary matrices $\Xi_0$ and $\Xi_1$
with $(\Xi_0)^4 = I$ and $(\Xi_1)^4 = I$
and using a suitable unitary matrix $W_0$, they are transformed as
\begin{eqnarray}
\Xi_0 \xrightarrow{W^{\dagger}_0 \Xi_0 W_0} \Xi_0^{\rm (D)},~~
\Xi_1 \xrightarrow{W^{\dagger}_0 \Xi_1 W_0} \tilde{\Xi}_1 \equiv V \Xi_1^{\rm (D)} V^{\dagger},
\label{Xi01}
\end{eqnarray}
where $V$ is a unitary matrix parameterized by
\begin{eqnarray}
V = e^{i\left(\xi^{a_{\scalebox{0.6}{$\displaystyle{ii}$}}} 
T^{a_{\scalebox{0.6}{$\displaystyle{ii}$}}}  
+ \xi^{a_{\scalebox{0.6}{$\displaystyle{-i~\!-i}$}}} 
T^{a_{\scalebox{0.6}{$\displaystyle{-i~\!-i}$}}}
+\xi^{a_{\scalebox{0.6}{$\displaystyle{i~\!-i}$}}} 
T^{a_{\scalebox{0.6}{$\displaystyle{i~\!-i}$}}}  
+ \xi^{a_{\scalebox{0.6}{$\displaystyle{-i~\!i}$}}} 
T^{a_{\scalebox{0.6}{$\displaystyle{-i~\!i}$}}}
+\xi^{a_{\scalebox{0.6}{$\displaystyle{-1~\!-1}$}}} 
T^{a_{\scalebox{0.6}{$\displaystyle{-1~\!-1}$}}}
\right)},
\label{V-Z4}
\end{eqnarray}
and $\Xi_0^{\rm (D)}$ and $\Xi_2^{\rm (D)}$ are
diagonal matrices expressed by
\begin{eqnarray}
&~& \mbox{diag} \Xi_0^{\rm (D)} = \left([1]_{p_1}, [1]_{p_2}, [i]_{p_3},
[i]_{p_4}, [-1]_{p_5}, [-1]_{p_6}, 
[-i]_{p_7}, [-i]_{p_8}\right),~~
\label{diag0-Z4} \\
&~& \mbox{diag} \Xi_1^{\rm (D)} = \left([1]_{p_1}, [-1]_{p_2}, [-i]_{p_3},
[i]_{p_4}, [-1]_{p_5}, [1]_{p_6}, [i]_{p_7}, [-i]_{p_8}\right).
\label{diag2-Z4}
\end{eqnarray}
Here, $[1]_{p_a}$, $[-1]_{p_a}$,  $[i]_{p_a}$ and $[-i]_{p_a}$
represent $1$, $-1$, $i$, and $-i$ for all $p_a$ elements,
and $0 \leq p_a \leq N$ ($a=1, \cdots, 8$).
The $T^{a_{\scalebox{0.6}{$\displaystyle{st}$}}}$  
are generators satisfy the relations:
\begin{eqnarray}
\Xi_0^{\rm (D)} T^{a_{\scalebox{0.6}{$\displaystyle{st}$}}}
= s T^{a_{\scalebox{0.6}{$\displaystyle{st}$}}} \Xi_0^{\rm (D)},~~
\Xi_1^{\rm (D)} T^{a_{\scalebox{0.6}{$\displaystyle{st}$}}}
= t T^{a_{\scalebox{0.6}{$\displaystyle{st}$}}} \Xi_1^{\rm (D)},
\label{Rel-Z4}
\end{eqnarray}
where $s$ and $t$ are $1$, $-1$, $i$, or $-i$.

Under a local gauge transformation,
$\Xi_0^{\rm (D)}$ and $\tilde{\Xi}_1$ change as
\begin{eqnarray}
{\Xi'}_0^{\rm (D)} = \varOmega(iz, -i\bar{z}) \Xi_0^{\rm (D)}
\varOmega^\dagger (z, \bar{z}),~~
\tilde{\Xi}'_1 = \varOmega(iz + 1, -i\bar{z}+1) \tilde{\Xi}_1 \varOmega^\dagger (z, \bar{z}),
\label{Xi'-Z4}
\end{eqnarray}
where $\varOmega$ is a gauge transformation function.
In a similar way as the cases of $T^2/Z_2$ and $T^2/Z_3$,
we find that local gauge transformations 
that make $\Theta_0^{\rm (D)}$ and $\tilde{\Theta}_1$
diagonal matrices simultaneously do not exist except for a special case.
Hence, diagonal representatives do not necessarily exist 
in equivalence classes of BC matrices on $T^2/Z_4$, either.

\subsection{$T^2/Z_6$}
\label{app-A4}

\begin{figure}[ht!]
\caption{Orbifold $T^2/Z_6$}
\label{FZ6}
\begin{center}
\unitlength 0.1in
\begin{picture}( 31.9000, 14.2000)(  6.8000,-25.2000)
%
\special{pn 13}%
\special{pa 2610 2390}%
\special{pa 3780 2390}%
\special{fp}%
\special{sh 1}%
\special{pa 3780 2390}%
\special{pa 3714 2370}%
\special{pa 3728 2390}%
\special{pa 3714 2410}%
\special{pa 3780 2390}%
\special{fp}%
%
\special{pn 13}%
\special{pa 2590 2380}%
\special{pa 774 1346}%
\special{fp}%
\special{sh 1}%
\special{pa 774 1346}%
\special{pa 822 1396}%
\special{pa 820 1372}%
\special{pa 842 1362}%
\special{pa 774 1346}%
\special{fp}%
%
\special{pn 8}%
\special{pa 790 1340}%
\special{pa 1980 1340}%
\special{dt 0.045}%
%
\special{pn 13}%
\special{sh 1}%
\special{ar 2620 2400 10 10 0  6.28318530717959E+0000}%
\special{sh 1}%
\special{ar 2610 2400 10 10 0  6.28318530717959E+0000}%
%
\special{pn 13}%
\special{sh 1}%
\special{ar 2280 1880 10 10 0  6.28318530717959E+0000}%
\special{sh 1}%
\special{ar 2270 1880 10 10 0  6.28318530717959E+0000}%
\put(24.4000,-26.0000){\makebox(0,0)[lb]{O}}%
\put(38.7000,-24.7000){\makebox(0,0)[lb]{$e_1$}}%
\put(6.8000,-12.7000){\makebox(0,0)[lb]{$e_2$}}%
\put(10.1000,-18.3000){\makebox(0,0)[lb]{$\frac{2e_2}{3}$}}%
\put(17.3000,-22.9000){\makebox(0,0)[lb]{$\frac{e_2}{3}$}}%
%
\special{pn 13}%
\special{sh 1}%
\special{ar 1710 1880 10 10 0  6.28318530717959E+0000}%
\special{sh 1}%
\special{ar 1700 1880 10 10 0  6.28318530717959E+0000}%
%
\special{pn 13}%
\special{sh 1}%
\special{ar 3170 2390 10 10 0  6.28318530717959E+0000}%
\special{sh 1}%
\special{ar 3160 2390 10 10 0  6.28318530717959E+0000}%
\put(31.3000,-26.9000){\makebox(0,0)[lb]{$\frac{e_1}{2}$}}%
\put(14.2000,-20.4000){\makebox(0,0)[lb]{$\frac{e_2}{2}$}}%
\put(19.9000,-17.7000){\makebox(0,0)[lb]{$\frac{e_1+e_2}{2}$}}%
%
\special{pn 13}%
\special{sh 1}%
\special{ar 2010 2050 10 10 0  6.28318530717959E+0000}%
\special{sh 1}%
\special{ar 2000 2050 10 10 0  6.28318530717959E+0000}%
%
\special{pn 13}%
\special{sh 1}%
\special{ar 1390 1700 10 10 0  6.28318530717959E+0000}%
\special{sh 1}%
\special{ar 1380 1700 10 10 0  6.28318530717959E+0000}%
%
\special{pn 8}%
\special{pa 3740 2340}%
\special{pa 1976 1348}%
\special{dt 0.045}%
\end{picture}%
\end{center}
\end{figure}
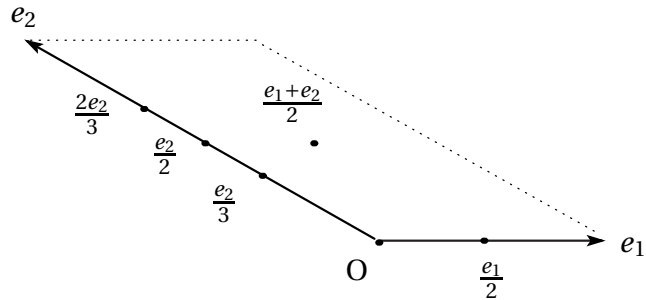

The orbifold $T^2/Z_6$ is obtained 
by identifying $z + e_1$, $z + e_2$, and $\varphi z$ with $z$.
Here, $T^2$ is constructed by the $G_2$ lattice
whose basis vectors are $e_1 = 1$ and $e_2 = (-3+i\sqrt{3})/2$,
and $\varphi = e^{\pi i/3}$.
The resultant space is depicted in Figure \ref{FZ6}.
The basis vectors are transformed as $\varphi e_1=2e_1+e_2$, 
$\varphi e_2=-3e_1-e_2$ under the $Z_6$ transformation
$z \to \varphi z$. 
There are one fixed point $z=0$ under the $Z_6$ transformation
$z \to \varphi z$, three fixed points
$z = 0$, $e_2/3$, and $2e_2/3$ under the $Z_3$ transformation 
$z \to \varphi^2 z = \omega z$
and four fixed points $z = 0$, $e_1/2$, $e_2/2$, and $(e_1+e_2)/2$ 
under the $Z_2$ transformation $z \to \varphi^3 z = - z$,
and, around these points, we define ten kinds of transformations:
\begin{eqnarray}
&~& R_0: z\rightarrow \varphi z, ~~
R_{10}: z\rightarrow \varphi^2 z, ~~
R_{11}: z\rightarrow \varphi^2 z+e_1+e_2, ~~
R_{12}: z\rightarrow \varphi^2 z+2e_1+2e_2, \nonumber \\
&~& R_{20}: z\rightarrow \varphi^3 z, ~~
R_{21}: z\rightarrow \varphi^3 z+e_1, ~~
R_{22}: z\rightarrow \varphi^3 z+e_2, ~~
R_{23}: z\rightarrow \varphi^3 z+e_1+e_2, \nonumber \\
&~& T_1: z\rightarrow z+e_1, ~~
T_2: z\rightarrow z+e_2,
\label{Z6-transf}
\end{eqnarray}
and they satisfy the relations:
\begin{eqnarray}
&~& R_0^6= I,~~ R_{10}^3= I,~~ R_{11}^3= I,~~ R_{12}^3= I,~~
R_{20}^2= I,~~ R_{21}^2=I,~~ R_{22}^2= I,~~ R_{23}^2=I,~~
\nonumber \\
&~& R_{11}=T_1 T_2 R_{10},~~ R_{12}=T_1^2 T_2^2 R_{10},~~
R_{21}=T_1 R_{20},~~ R_{22}=T_2 R_{20},~~
\nonumber \\
&~& R_{23}= T_1 T_2 R_{20}=R_{21} R_{20} R_{22}
=R_{22} R_{20} R_{21}=R_{11} R_0,
\nonumber \\
&~& R_{10}=R_0^2,~~ R_{20}=R_0^3,~~ T_1 T_2=T_2 T_1,~~
T_2 = R_0^2 T_1 R_0 T_1 R_0^3,
\nonumber \\
&~& (R_0 R_{10})^4 = I,~~ (R_0 R_{11})^4 = I,~~ (R_0 R_{12})^4 = I,~~
\nonumber \\
&~& (R_0 R_{20})^3 = I,~~ (R_0 R_{21})^3 = I,~~
(R_0 R_{22})^3 = I,~~ (R_0 R_{23})^3 = I.
\label{Z6-relations}
\end{eqnarray}
We denote the BC matrix for the $Z_6$ transformation
$R_0:z \to e^{\pi i/3} z$ as $\Phi$, and {\it other representation
matrices are determined uniquely, using the relations {\rm (\ref{Z6-relations})},
if $\Phi$ is given.}
Because $\Phi$ is a unitary matrix, it can be diagonalized by 
a suitable unitary transformation.
Hence, each equivalence class of BC matrices on $T^2/Z_6$
owns a diagonal representative.

\section{Conclusions and discussions}

We have studied diagonal representatives 
of BC matrices on the orbifolds $S^1/Z_2$ and $T^2/Z_m$ ($m=2, 3, 4, 6$).
We have given an alternative proof of the existence of diagonal representatives 
in each equivalent class of BC matrices
on $S^1/Z_2$, using a matrix exponential representation, 
and shown that diagonal representatives do not necessarily exist on $T^2/Z_2$,
$T^2/Z_3$, and $T^2/Z_4$.
We have found that there is a diagonal representative in each equivalence class
on $T^2/Z_6$, because its BCs are determined by a single unitary matrix.

In the presence of equivalence classes of BC matrices 
without diagonal representatives
on $T^2/Z_2$, $T^2/Z_3$, and $T^2/Z_4$,
it demonstrates that constructions of realistic models and 
phenomenological studies have not been completed.
In particular,
it would be interesting to investigate 
model-buildings concerning grand unification, gauge-Higgs unification,
grand gauge-Higgs unification~\cite{KT&Y,Y},
and/or family unification based on non-trivial BC matrices.
Furthermore, there remains the arbitrariness problem
of which type of BC matrices should be chosen 
without relying on phenomenological information.
It would be a challenging problem to find a mechanism or principle
that determines BCs of fields.

\appendix

\section{Generators and matrix representation}

For $P_0^{\rm (D)}$ and $P_1^{\rm (D)}$ defined by Eqs.~(\ref{diag0}) and (\ref{diag1}),
using the relations (\ref{T++}), (\ref{T+-}), (\ref{T-+}), and (\ref{T--}),
the generators of $U(N)$ are classified into 4 types 
$\{T^{a_{\scalebox{0.6}{$\displaystyle{++}$}}},
T^{a_{\scalebox{0.6}{$\displaystyle{+-}$}}},
T^{a_{\scalebox{0.6}{$\displaystyle{-+}$}}},
T^{a_{\scalebox{0.6}{$\displaystyle{--}$}}}\}$
described by
\begin{eqnarray}
&~& T^{a_{\scalebox{0.6}{$\displaystyle{++}$}}}
= \left(
\begin{array}{cccc}
\star & 0 & 0 & 0 \\
0 & \star & 0 & 0 \\
0 & 0 & \star & 0 \\
0 & 0 & 0 & \star  
\end{array} 
\right),~~
T^{a_{\scalebox{0.6}{$\displaystyle{+-}$}}}
= \left(
\begin{array}{cccc}
0 & \star & 0 & 0 \\
\star & 0 & 0 & 0 \\
0 & 0 & 0 & \star \\
0 & 0 & \star & 0
\end{array} 
\right),
\label{T+}\\
&~& T^{a_{\scalebox{0.6}{$\displaystyle{-+}$}}}
= \left(
\begin{array}{cccc}
0 & 0 & \star & 0 \\
0 & 0 & 0 & \star \\
\star & 0 & 0 & 0 \\
0 & \star & 0 & 0 \\
\end{array} 
\right),~~
T^{a_{\scalebox{0.6}{$\displaystyle{--}$}}}
= \left(
\begin{array}{cccc}
0 & 0 & 0 & \star \\
0 & 0 & \star & 0 \\
0 & \star & 0 & 0 \\
\star & 0 & 0 & 0 \\
\end{array} 
\right),
\label{T-}
\end{eqnarray}
where $\star$ stands for sub-matrices with non-zero elements,
and $0$ is a sub-matrix all of whose entries are zero (a null sub-matrix).
The numbers of $T^{a_{\scalebox{0.6}{$\displaystyle{++}$}}}$,
$T^{a_{\scalebox{0.6}{$\displaystyle{+-}$}}}$,
$T^{a_{\scalebox{0.6}{$\displaystyle{-+}$}}}$,
and $T^{a_{\scalebox{0.6}{$\displaystyle{--}$}}}$
are $p^2 + q^2 + r^2 + s^2$, $2(pq+rs)$, $2(pr+qs)$, and $2(ps+qr)$, respectively.
Here, $s = N-p-q-r$.
The total number of generators is $N^2$.

The commutation relations among those Lie algebras are given by
\begin{eqnarray}
&~& \left[T^{a_{\scalebox{0.6}{$\displaystyle{++}$}}},
T^{b_{\scalebox{0.6}{$\displaystyle{st}$}}}\right] 
= i f_{{a_{\scalebox{0.6}{$\displaystyle{++}$}}}{b_{\scalebox{0.6}{$\displaystyle{st}$}}}
{c_{\scalebox{0.6}{$\displaystyle{st}$}}}}T^{c_{\scalebox{0.6}{$\displaystyle{st}$}}},~~
\left[T^{a_{\scalebox{0.6}{$\displaystyle{st}$}}},
T^{b_{\scalebox{0.6}{$\displaystyle{st}$}}}\right] 
= i f_{{a_{\scalebox{0.6}{$\displaystyle{st}$}}}{b_{\scalebox{0.6}{$\displaystyle{st}$}}}
{c_{\scalebox{0.6}{$\displaystyle{++}$}}}}T^{c_{\scalebox{0.6}{$\displaystyle{++}$}}},~~
\nonumber \\
&~& \left[T^{a_{\scalebox{0.6}{$\displaystyle{+-}$}}},
T^{b_{\scalebox{0.6}{$\displaystyle{--}$}}}\right] 
= i f_{{a_{\scalebox{0.6}{$\displaystyle{+-}$}}}{b_{\scalebox{0.6}{$\displaystyle{--}$}}}
{c_{\scalebox{0.6}{$\displaystyle{-+}$}}}}T^{c_{\scalebox{0.6}{$\displaystyle{-+}$}}},~~
\left[T^{a_{\scalebox{0.6}{$\displaystyle{-+}$}}},
T^{b_{\scalebox{0.6}{$\displaystyle{--}$}}}\right] 
= i f_{{a_{\scalebox{0.6}{$\displaystyle{-+}$}}}{b_{\scalebox{0.6}{$\displaystyle{--}$}}}
{c_{\scalebox{0.6}{$\displaystyle{+-}$}}}}T^{c_{\scalebox{0.6}{$\displaystyle{+-}$}}},~~
\nonumber \\
&~& \left[T^{a_{\scalebox{0.6}{$\displaystyle{+-}$}}},
T^{b_{\scalebox{0.6}{$\displaystyle{-+}$}}}\right] 
= i f_{{a_{\scalebox{0.6}{$\displaystyle{+-}$}}}{b_{\scalebox{0.6}{$\displaystyle{-+}$}}}
{c_{\scalebox{0.6}{$\displaystyle{--}$}}}}T^{c_{\scalebox{0.6}{$\displaystyle{--}$}}},
\label{T-Rel}
\end{eqnarray}
where 
$f_{{a_{\scalebox{0.6}{$\displaystyle{++}$}}}{b_{\scalebox{0.6}{$\displaystyle{st}$}}}
{c_{\scalebox{0.6}{$\displaystyle{st}$}}}}$,
$f_{{a_{\scalebox{0.6}{$\displaystyle{st}$}}}{b_{\scalebox{0.6}{$\displaystyle{st}$}}}
{c_{\scalebox{0.6}{$\displaystyle{++}$}}}}$,
$f_{{a_{\scalebox{0.6}{$\displaystyle{+-}$}}}{b_{\scalebox{0.6}{$\displaystyle{--}$}}}
{c_{\scalebox{0.6}{$\displaystyle{-+}$}}}}$,
$f_{{a_{\scalebox{0.6}{$\displaystyle{-+}$}}}{b_{\scalebox{0.6}{$\displaystyle{--}$}}}
{c_{\scalebox{0.6}{$\displaystyle{+-}$}}}}$, and
$f_{{a_{\scalebox{0.6}{$\displaystyle{+-}$}}}{b_{\scalebox{0.6}{$\displaystyle{-+}$}}}
{c_{\scalebox{0.6}{$\displaystyle{--}$}}}}$
are structure constants, and
$s$ and $t$ are $+$ or $-$.

Using a matrix representation, an arbitrary unitary matrix $W$ 
and its hermitian conjugated one $W^{\dagger}$
can be written by 
\begin{eqnarray}
W = \left(
\begin{array}{cccc}
A_1 & B_1 & C_1 & D_1 \\
B_2 & A_2 & D_2 & C_2 \\
C_3 & D_3 & A_3 & B_3 \\
D_4 & C_4 & B_4 & A_4  
\end{array} 
\right),~~
W^{\dagger} = \left(
\begin{array}{cccc}
A^{\dagger}_1 & B^{\dagger}_2 & C^{\dagger}_3 & D^{\dagger}_4 \\
B^{\dagger}_1 & A^{\dagger}_2 & D^{\dagger}_3 & C^{\dagger}_4 \\
C^{\dagger}_1 & D^{\dagger}_2 & A^{\dagger}_3 & B^{\dagger}_4 \\
D^{\dagger}_1 & C^{\dagger}_2 & B^{\dagger}_3 & A^{\dagger}_4  
\end{array} 
\right),
\label{W-matrix}
\end{eqnarray}
where $A_{k}$, $B_{k}$, $C_{k}$, and $D_{k}$ ($k=1,2,3,4$)
are sub-matrices that satisfy the relations from unitarity:
\begin{eqnarray}
&~& A_k A^{\dagger}_k + B_k B^{\dagger}_k + C_k C^{\dagger}_k + D_k D^{\dagger}_k 
= I,
\label{U-rel1}\\
&~& A_1 B^{\dagger}_2 + B_1 A^{\dagger}_2 + C_1 D^{\dagger}_2 + D_1 C^{\dagger}_2 = 0,~~
A_1 C^{\dagger}_3 + B_1 D^{\dagger}_3 + C_1 A^{\dagger}_3 + D_1 B^{\dagger}_3 = 0,
\label{U-rel2}\\
&~& A_1 D^{\dagger}_4 + B_1 C^{\dagger}_4 + C_1 B^{\dagger}_4 + D_1 A^{\dagger}_4 = 0,~~
B_2 C^{\dagger}_3 + A_2 D^{\dagger}_3 + D_2 A^{\dagger}_3 + C_2 B^{\dagger}_3 = 0,
\label{U-rel3}\\
&~& B_2 D^{\dagger}_4 + A_2 C^{\dagger}_4 + D_2 B^{\dagger}_4 + C_2 A^{\dagger}_4 = 0,~~
C_3 D^{\dagger}_4 + D_3 C^{\dagger}_4 + A_3 B^{\dagger}_4 + B_3 A^{\dagger}_4 = 0,
\label{U-rel4}\\
&~& A^{\dagger}_1 A_1 + B^{\dagger}_2 B_2 + C^{\dagger}_3 C_3 + D^{\dagger}_4 D_4 = I,~~
A^{\dagger}_2 A_2 + B^{\dagger}_1 B_1 + C^{\dagger}_4 C_4 + D^{\dagger}_3 D_3 = I,
\label{U-rel5}\\
&~& A^{\dagger}_3 A_3 + B^{\dagger}_4 B_4 + C^{\dagger}_1 C_1 + D^{\dagger}_2 D_2 = I,~~
A^{\dagger}_4 A_4 + B^{\dagger}_3 B_3 + C^{\dagger}_2 C_2 + D^{\dagger}_1 D_1 = I,
\label{U-rel6}\\
&~& A^{\dagger}_1 B_1 + B^{\dagger}_2 A_2 + C^{\dagger}_3 D_3 + D^{\dagger}_4 C_4 = 0,~~
A^{\dagger}_1 C_1 + B^{\dagger}_2 D_2 + C^{\dagger}_3 A_3 + D^{\dagger}_4 B_4 = 0,
\label{U-rel7}\\
&~& A^{\dagger}_1 D_1 + B^{\dagger}_2 C_2 + C^{\dagger}_3 B_3 + D^{\dagger}_4 A_4 = 0,~~
B^{\dagger}_1 C_1 + A^{\dagger}_2 D_2 + D^{\dagger}_3 A_3 + C^{\dagger}_4 B_4 = 0,
\label{U-rel8}\\
&~& B^{\dagger}_1 D_1 + A^{\dagger}_2 C_2 + D^{\dagger}_3 B_3 + C^{\dagger}_4 A_4 = 0,~~
C^{\dagger}_1 D_1 + D^{\dagger}_2 C_2 + A^{\dagger}_3 B_3 + B^{\dagger}_4 A_4 = 0.
\label{U-rel9}
\end{eqnarray}
Here, $I$ and $0$ represent a unit sub-matrix and a null sub-matrix, respectively. 

By using three unitary matrices:
\begin{eqnarray}
U_1 = \left(
\begin{array}{cccc}
a_1 & b_1 & 0 & 0 \\
b_2 & a_2 & 0 & 0 \\
0 & 0 & a_3 & b_3 \\
0 & 0 & b_4 & a_4  
\end{array} 
\right),~~
U_2 = \left(
\begin{array}{cccc}
e_1 & 0 & 0 & d_1 \\
0 & e_2 & d_2 & 0 \\
0 & d_3 & e_3 & 0 \\
d_4 & 0 & 0 & e_4  
\end{array} 
\right),
U_3 = \left(
\begin{array}{cccc}
f_1 & 0 & c_1 & 0 \\
0 & f_2 & 0 & c_2 \\
c_3 & 0 & f_3 & 0 \\
0 & c_4 & 0 & f_4  
\end{array} 
\right),
\label{Us}
\end{eqnarray}
a generic $N \times N$ unitary matrix $W$ can be parametrized as
\begin{eqnarray}
\hspace{-1cm}&~& W = U_1 U_2 U_3 
\nonumber \\
\hspace{-1cm}&~& ~~~~~ = \left(
\begin{array}{cccc}
a_1 e_1 f_1 + b_1 d_2 c_3 & a_1 d_1 c_4 + b_1 e_2 f_2 & 
a_1 e_1 c_1 + b_1 d_2 f_3 & a_1 d_1 f_4 + b_1 e_2 c_2 \\
b_2 e_1 f_1 + a_2 d_2 c_3 & b_2 d_1 c_4 + a_2 e_2 f_2 & 
b_2 e_1 c_1 + a_2 d_2 f_3 & b_2 d_1 f_4 + a_2 e_2 c_2 \\
a_3 e_3 c_3 + b_3 d_4 f_1 & a_3 d_3 f_2 + b_3 e_4 c_4 & 
a_3 e_3 f_3 + b_3 d_4 c_1 & a_3 d_3 c_2 + b_3 e_4 f_4 \\
b_4 e_3 c_3 + a_4 d_4 f_1 & b_4 d_3 f_2 + a_4 e_4 c_4 & 
b_4 e_3 f_3 + a_4 d_4 c_1 & b_4 d_3 c_2 + a_4 e_4 f_4
\end{array} 
\right),
\label{W-Vs}
\end{eqnarray}
where $a_{k}$, $b_{k}$, $c_{k}$, and $d_{k}$ are sub-matrices that satisfy 
relations from unitarity.
Using a matrix representation,
unitary matrices given by the matrix exponential representation such as
$e^{i\left(\xi^{a_{\scalebox{0.6}{$\displaystyle{++}$}}} 
T^{a_{\scalebox{0.6}{$\displaystyle{++}$}}}  
+ \xi^{a_{\scalebox{0.6}{$\displaystyle{+-}$}}} 
T^{a_{\scalebox{0.6}{$\displaystyle{+-}$}}}\right)}$,
$e^{i \xi^{a_{\scalebox{0.6}{$\displaystyle{--}$}}} 
T^{a_{\scalebox{0.6}{$\displaystyle{--}$}}}}$,
and 
$e^{i\left(\eta^{a_{\scalebox{0.6}{$\displaystyle{++}$}}} 
T^{a_{\scalebox{0.6}{$\displaystyle{++}$}}}  
+ \xi^{a_{\scalebox{0.6}{$\displaystyle{-+}$}}}
T^{a_{\scalebox{0.6}{$\displaystyle{-+}$}}}\right)}$
are parametrized by $U_1$, $U_2$, and $U_3$, respectively.
Hence, $W$ is represented by Eq.~(\ref{W}).
The $W$ contains $p^2 + q^2 + r^2 + s^2$ redundant parameters
$\eta^{a_{\scalebox{0.6}{$\displaystyle{++}$}}}$.
We add them to make $V_3$ represent arbitrary elements of 
a subgroup of $U(N)$.

\section*{Acknowledgments}
The authors thank Prof. Yamashita for valuable discussions.
This work was supported in part by scientific grants 
from the Ministry of Education, Culture,
Sports, Science and Technology under Grant No.~17K05413 (YK).

\end{document}